\documentstyle[aps,epsfig]{revtex}
\def\bea{\begin{eqnarray}}
\def\beann{\begin{eqnarray*}}
\def\beq{\begin{equation}}
\def\eea{\end{eqnarray}}
\def\eeann{\end{eqnarray*}}
\def\eeq{\end{equation}}
\def\nn{\nonumber}
\begin{document}
\title{Excluded Volume Effects in the Quark Meson Coupling Model}
\author{P. K. Panda$^1$, M. E. Bracco$^2$,
M. Chiapparini$^2$, E. Conte$^2$ and G. Krein$^1$}
\address{$^1$ Instituto de F\'{\i}sica Te\'orica, Universidade 
Estadual Paulista\\
Rua Pamplona 145, 01405-900 S\~ao Paulo, SP, Brasil}
\address{$^2$ Instituto de F\'\i sica, Universidade do Estado 
do Rio de Janeiro \\
Rua S\~ao Francsico Xavier 524, 20559-900, Rio de Janeiro, RJ, Brasil}
\maketitle
\begin{abstract}
Excluded volume effects are incorporated in the quark meson coupling model 
to take into account in a phenomenological way the hard core repulsion 
of the nuclear force. The formalism employed is thermodynamically consistent 
and does not violate causality. The effects of the excluded volume on in-medium
nucleon properties and the nuclear matter equation of state are 
investigated as a function of the size of the hard core.
It is found that in-medium nucleon properties 
are not altered significantly by the excluded volume, even for large hard 
core radii, and the equation of state becomes stiffer as the size of 
the hard core increases. 
\end{abstract}
\pacs{12.39.Ba, 21.65.+f, 24.85.+p}

\section{Introduction}
The study of the properties of high density and high temperature
hadronic matter is of interest for  understanding a wide 
range of phenomena associated with superdense stars~\cite{HP} and
relativistic heavy-ion collisions~\cite{HM}. One of the open 
questions in this subject is the correct identification of the 
appropriate degrees of freedom to describe the different phases 
of hadronic matter. Although this question will eventually be 
answered with first-principles calculations with the fundamental 
theory of the strong interactions quantum chromodynamics (QCD), 
most probably through lattice QCD simulations, presently one is still 
far from this goal and, in order to make progress, one must rely on model
calculations and make use of the scarce experimental information 
available. For matter at zero temperature and density close to the 
saturation density of nuclear matter, experiment seems to indicate 
that the relevant degrees of freedom are the baryons and mesons. 
There is a long and successful history of calculations using 
models based on baryonic and mesonic degrees of freedom, such as 
potential models~\cite{FW,WFF} and relativistic field-theoretical 
models, generically known as quantum hadrodynamics (QHD)~\cite{WalSer,Eff}. 
At densities several times larger than the saturation density and/or 
high temperatures, one expects a phase of deconfined matter whose 
properties are determined by the internal degrees of freedom of the 
hadrons. Early studies of deconfined matter~\cite{bagEOS} used the 
MIT bag model~\cite{MITbag}, in which the relevant degrees of 
freedom are quarks and gluons confined by the vacuum 
pressure. On the other hand, at high densities, but not asymptotically 
higher than the saturation density, the situation seems to be very 
complicated. The complication arises because of the possibility of 
simultaneous presence of hadrons and deconfined quarks and gluons in 
the system. Not much progress has been possible in this direction 
because of the necessity of a model able to describe composites and 
constituents at the same footing - a step towards this direction is 
the formalism developed in Ref.~\cite{FT}.

An important step towards the formulation of a model to describe
the different phases of hadronic matter in terms of explicit 
quark-gluon degrees of freedom is the quark-meson coupling (QMC)
model, proposed by Guichon~\cite{guichon} some time ago and 
extensively investigated by Saito, Thomas and collaborators
~\cite{ST} - see also Ref.~\cite{earlier} for related studies. 
Matter at low density and temperature is described as a system of 
nonoverlapping bags interacting through effective scalar- and 
vector-meson degrees of freedom, very much in the same way as in 
QHD~\cite{WalSer,Eff}. The crucial difference is that in the QMC, 
the effective mesons couple directly to the quarks in the interior
of the baryons, with the consequence that the effective 
baryon-meson coupling constants become density dependent. 
In addition, hadronic sizes are explicitly incorporated through 
form factors calculable within the underlying quark 
model~\cite{ffactors}. At very high density and/or temperature, 
baryons and mesons dissolve and the entire system of
deconfined matter, composed by quarks and gluons, becomes 
confined within a single MIT bag~\cite{bagEOS}.

The fact that the same underlying quark model is used at 
different phases of hadronic matter makes the model very attractive
conceptually. Many applications and extensions of the model 
have been made in the last years - see 
Refs.~\cite{ffactors,bstar,recent,stars,LTTWS,temp,qmcstar} and references 
therein. Of~particular interest for the phenomenology of finite 
nuclei was the introduction of a density-dependent bag constant 
by Jin and Jennings~\cite{bstar}. These authors postulated
different density dependencies for the bag constant, in a way 
that the bag constant decreases as the nuclear density increases. 
One consequence of this is that large values for the scalar and 
vector mean fields at the saturation density are obtained, leading
to spin-orbit splittings of the single-particle levels of finite 
nuclei that are in better agreement with experiment than those 
obtained with a density independent bag constant. Another 
consequence of a smaller bag constant in medium is that the bag 
radius becomes considerably larger than in free 
space~\cite{bstar,stars}. At saturation density the nucleon radius
increases 25\% and at densities three times higher the radius can
increase as much as 50\%. For higher densities the increase of radius
is even more dramatic. The consequences of changing the bag constant 
for nucleon sizes were investigated by Lu et al.~\cite{LTTWS}. 

A large increase of the bag radius naturally raises the question 
about the validity of the nonoverlapping bag picture that underlies 
calculations of nuclear matter properties with the model. At normal 
nuclear matter densities, the average distance between nucleons is 
of the order of 1.8~fm. Therefore, for densities larger than the 
normal density and bag radii larger than 1~fm there is a large 
probability that the bags overlap significantly. However, before 
concluding that the picture of independent nucleons breaks down, it is 
important to recall that short-distance correlations are left out 
in a mean field calculation. These correlations are induced by the 
combined effects of the Pauli exclusion principle between identical 
nucleons and the hard-core of the nucleon-nucleon interaction that 
forbids scattering into occupied levels. The success of the independent 
particle model of the nuclear shell model is due to the small size of 
hard core and the Pauli Principle that lead to a ``healing" distance of 
the two nucleon relative wave function that is smaller than the average 
distance between nucleons in medium~\cite{GWW}. In a model like the QMC, where 
the finite size of the nucleons is made explicit through a bag structure, 
the incorporation 
of this physics in the many-body dynamics is an interesting new development. 
In the present paper we address this in a phenomenological way through an 
excluded volume approach. The prescription we use was developed in
Ref.~\cite{Goren1} for ideal gases and further extended to relativistic 
field-theoretic models as QHD in Ref.~\cite{Goren2}. In this approach, 
matter in the hadronic phase is described by nonoverlapping rigid 
spheres, but when the density of matter is such that the relative 
distance between two spheres becomes smaller than the diameter of 
the spheres, the excluded volume effect introduces an effective
repulsion that mimics the hard-core repulsion of the nucleon-nucleon 
interaction. 

Of course, at very high density the description of hadronic matter in
terms of nonoverlapping bags should break down. In a purely geometrical 
view, one has the picture that once the relative distance between two 
bags becomes much smaller than the diameter of a bag, quarks and gluons 
start percolate and individual bags loose their identity. The density 
for which this starts to happen is presently unknown within QCD.
In this respect, it is important not 
to confuse the bag radius with the radius of the hard-core of the 
nucleon-nucleon force. Model studies~\cite{goldman} indicate 
that when the two nucleons start to overlap, medium-range forces are 
generated from the distortion of the quark distribution. The short-range
repulsion, on the other hand, is due to the combined effect of the 
one-gluon exchange - mainly due to its spin-spin component - and the 
Pauli exclusion principle between quarks of different nucleons that
becomes efficient when the overlap of the two-nucleon wave functions 
is complete. Although the described scenario might well not be 
ultimately confirmed by a full QCD calculation, it seems however 
clear that the two radii, the radius of the MIT bag and the radius 
of the hard-core of the nucleon-nucleon interaction, are of different 
sizes and have different origin in the physics of hadron structure. 
In this sense, the volume of the excluded volume will be taken to be 
smaller and unrelated to the radius of the underlying MIT bag.

The excluded volume approach we use is thermodynamically consistent.  
Although the prescription can be extended to take into account Lorentz 
contraction of the bags~\cite{Goren3}, in this initial exploratory 
investigation we use hard-sphere bags, since a complete calculation 
would lead to massive numerical calculations. 
In addition, as indicated by the investigations in Ref.~\cite{Goren3}, 
the effect of Lorentz contraction is most important for light particles 
like pions. However, as we will explicitly show, the approach does not 
lead to violations of causality for the density range where the model 
makes sense. 

The paper is organized as follows. In Section~\ref{rev} we present
a short review of the excluded volume prescription of 
Ref.~\cite{Goren2} and implement it to the QMC model. Numerical 
results are presented in Section~\ref{numer} and our Conclusions 
and Perspectives are discussed in Section~\ref{concl}.  

\section{Excluded Volume in the QMC Model}
\label{rev}

Initially, for completeness and in order to make the paper selfcontained, 
we briefly recapitulate the excluded volume prescription of Ref.~\cite{Goren2}. 
Let us start with the ideal gas of one particle species with temperature $T$, 
chemical potential $\mu$ and volume $V$. The pressure is related to the grand 
partition function ${\cal Z}$ as
\begin{equation}
P(T,\mu)=\lim_{V\rightarrow\infty}T~\frac{ln ~{\cal Z}(T,\mu,V)}{V},
\end{equation}
with ${\cal Z}$ defined as
\begin{equation}
{\cal Z}(T,\mu,V)=\sum_{N=0}^{\infty} e^{-\mu N/T} Z (T,N,V),
\label{Zcan}
\end{equation}
where $Z$ is the canonical partition function. The authors in Ref.~\cite{Goren2}
included the excluded volume effect starting from the canonical partition 
function as
\begin{equation}
Z^{excl} (T,N,V)=Z (T,N,V-v_0 N)\Theta(V-v_0 N).
\label{Zexcl}
\end{equation}
This ansatz is motivated by considering the volume $V$ for a system of
$N$ particles is reduced to an effective volume, $V-v_0 N$ where $v_0$ 
is the volume of a particle. In an hadronic gas, $v_0$ can be interpreted 
as the region excluded by the hard core of the nucleon-nucleon interaction.
For a spherical region, $v_0=4\pi r^3/3$ with $r$ the hard-core radius. 
Using Eq.~(\ref{Zexcl}) into Eq.~(\ref{Zcan}), the grand partition function 
becomes
\begin{equation}
{\cal Z}^{excl}(T,\mu,V)=\sum_{N=0}^{\infty} e^{-\mu N/T}
Z (T,N,V-v_0 N)\Theta(V-v_0 N).
\label{GP}
\end{equation}
There is a difficulty for evaluation of the sum over $N$ particles in this
equation because of the dependence of the available volume on the varying 
number of particles N because $Z (T,N,V-v_0 N)$ does not factor as a 
product as in the case of an $N$-independent volume. To overcome this 
difficulty the authors in Ref.~\cite{Goren2} have performed a Laplace 
transformation on the variable V in Eq.~(\ref{GP}) as
\begin{equation}
\tilde {\cal Z}^{excl}(T,\mu,\xi) = \int_0^\infty dV 
e^{-\xi V}{\cal Z}^{excl}(T,\mu,V).
\end{equation}
Using Eq.~(\ref{GP}) in this, and making the change of variable 
\begin{equation}
V = x + v_0 N,
\end{equation}
one obtains
\begin{equation}
\tilde {\cal Z}^{excl}(T,\mu,\xi) = 
\int_0^\infty d x  \, e^{-\xi x } {\cal Z}(T,\tilde \mu, x) = 
\tilde {\cal Z}^{excl}(T,\tilde \mu,\xi),
\end{equation}
where $\tilde \mu=\mu - v_0 T \xi$. Now the integrand in this is 
factorizable (for the present case of independent particles) and the
sum over $N$ can be implemented. It is a simple exercise~\cite{Goren1} 
to show that the pressure of the system is given as~\cite{Goren2}
\begin{equation}
P(T,\mu)=P^\prime(T,\tilde\mu),
\label{ev1}
\end{equation}
with 
\begin{equation}
\tilde\mu =\mu- v_0 \, P(T,\mu).
\label{ev2}
\end{equation} 
The meaning of Eq.~(\ref{ev1}) is that the pressure of the system with
excluded volume with chemical potential $\mu$, $P(T,\mu)$,  
is equal to the pressure of a system without excluded volume but with
an effective chemical potential $\tilde \mu = \mu- v_0 P(T,\mu)$,  
denoted by $P^\prime(T,\tilde\mu)$. Note that once the expression for
$P^\prime(T,\tilde\mu)$ is known, the pressure of the 
system is given by an implicit function.

The baryon density, the entropy density and the energy density
for the system are given by the usual thermodynamical expressions
\begin{eqnarray}
\rho(T,\mu) &\equiv& \Bigg(\frac{\partial P}{\partial\mu}\Bigg)_T =
\frac{\rho^\prime(T, \tilde\mu)}{1+ v_0
\rho^\prime(T, \tilde\mu)},
\label{ev3}
\\
S(T,\mu) &\equiv& \Bigg(\frac{\partial P}{\partial T}\Bigg)_{\mu} =
\frac{S^\prime(T, \tilde\mu)}{1+ v_0\rho^\prime(T,\tilde\mu)},
\label{ev4}
\\
\epsilon(T,\mu) &\equiv& TS -P +\mu\rho =
\frac{\epsilon^\prime(T, \tilde\mu)}{1+ v_0 \rho^\prime(T, \tilde\mu)}.
\label{ev5}
\end{eqnarray}
These relations define a thermodynamical consistent formalism, since the 
fundamental thermodynamical relations are fulfilled. 

Next we apply this formalism to the QMC model for nuclear matter at zero 
temperature~\cite{guichon,ST}. In the QMC model, the nucleon in nuclear matter 
is assumed to be described by a static MIT bag in which quarks interact with 
scalar $\sigma_0$ and vector $\omega_0$ mean mesonic fields. The mesonic fields 
are meant to represent effective degrees of freedom, not necessarily identified
with real mesons. Therefore, since the introduction of the excluded volume is 
to represent the hard core nucleon-nucleon interaction,
Eqs.~(\ref{ev1})-(\ref{ev5}) will be applied to the baryons only. The same 
prescription has been used in the application of the formalism to QHD in 
Ref.~\cite{Goren2}.  

In the QMC model, the pressure and energy density receive contributions from 
baryons and mesons and are given as
\begin{eqnarray}
P &=& P_B -  \frac{1}{2} m^2_\sigma \sigma^2_0 
+ \frac{1}{2} m^2_\omega \omega^2_0 ,\\
\epsilon  &=& \epsilon_B +  \frac{1}{2} m^2_\sigma \sigma^2_0 
+ \frac{1}{2} m^2_\omega \omega^2_0 ,
\end{eqnarray}
where $P_B$ and $\epsilon_B$ are the baryon contributions. As said above, the
excluded volume will be applied only to the baryonic contributions and, at zero 
temperature, one needs to consider only Eqs.~(\ref{ev3}) and (\ref{ev5}),
\begin{eqnarray}
\rho & = & \frac{\rho^\prime}{1+ v_0
\rho^\prime},
\label{ev3-0}
\\
\epsilon_B &=& \frac{\epsilon^\prime_B}{1+ v_0 \rho^\prime}.
\label{ev5-0}
\end{eqnarray}
For practical calculations, it is convenient to parameterize $\rho^\prime$ in
terms of a $k_F$ according to 
\beq
\rho^\prime = \frac{\gamma }{6\pi^2}\,k^3_F.
\label{defk_F}
\eeq
This allows to write the QMC expressions for $P^\prime_B$ and $\epsilon^\prime_B$
as
\bea
P^\prime_B &=& \frac{1}{3}\frac{\gamma}{2\pi^2}
\Bigg[\frac{1}{4}k_F^3{\sqrt{k_F^2+{M^*}^2 }} - 
\frac{3}{8}{M^*}^2 k_F {\sqrt{k_F^2+{M^*}^2}} \nn\\
&& + \; \frac{3}{8}{M^*}^4 \ln \left( \frac{k_F+
{\sqrt{k_F^2+{M^*}^2}}}{M^*} \right)\Bigg], \\                  
\label{Pprime}
\epsilon^\prime_B &=& \rho^\prime {\sqrt{k^2_F+{M^*}^2 }}-P^\prime_B.
\label{eprime}
\eea
In these, $M^*$ is the in medium nucleon mass calculated with the MIT bag. Its 
value is determined by solving the MIT bag equations for quarks coupled to the
mean fields $\sigma_0$ and $\omega_0$. In order to completely determine 
$M^*$, and therefore the nuclear matter properties Eqs.~(\ref{ev3-0}) and 
(\ref{ev5-0}), one needs $\sigma_0$ and $\omega_0$. The scalar mean field is 
determined selfconsistently from the minimization condition at density $\rho$:
\begin{equation}
\frac{\partial \epsilon}{\partial \sigma_0} = 0 ,
\label{sig}
\end{equation}
which leads to
\beq
\sigma_0 = \frac{1}{1 + v_0 \rho^{\prime}} \frac{\Sigma(\sigma_0)}{m^2_\sigma}, 
\label{expl-sig}
\eeq
with
\beq
\Sigma(\sigma_0) = - \frac{1}{\pi^2} \, \frac{\partial M^*}{\partial \sigma_0}
\left[ k_F \sqrt{k^2_F + M^{* 2}} 
- M^{* 2} \ln \left( \frac{k_F + \sqrt{k^2_F + M^{* 2}}}{M^*} \right) \right].
\label{Sigma}
\eeq
The vector mean field $\omega_0$ is obtained from its equation of motion 
as
\begin{equation}
\omega_0 = \frac{3 g^q_\omega}{m_\omega^2}\,\rho .
\end{equation}

Solution of Eq.~(\ref{expl-sig}) proceeds as follows. For a given $\rho$, 
we use Eq.~(\ref{ev3-0}) to obtain $\rho^\prime$, and form this 
$\rho^\prime$ we obtain $k_F$ of Eq.~(\ref{defk_F}). The derivative 
$\partial M^*/\partial \sigma_0$ can be done explicitly, 
$\Sigma(\sigma_0)$ is then known, and the transcendental equation for 
$\sigma_0$ is easily solved numerically. The results are presented 
in the next section.

\section{Results and Discussions}
\label{numer}

We start fixing the free-space bag properties. We use zero quark masses only 
and use two values for the bag radius, $R = 0.6$~fm and $R = 0.8$~fm. There are 
two unknowns, $z_0$ and the bag constant $B$. These are obtained as usual by 
fitting the nucleon mass $M=939$~MeV and enforcing the stability condition 
for the bag. The values obtained for $z_0$ and $B$ are displayed in Table~1. 

Next we proceed to nuclear matter properties. We will consider two versions
of the model. In the first one, the bag constant $B$ is fixed at its vacuum 
value, and in the second one the bag constant changes accordingly to the original 
Jin and Jennings\cite{bstar} ansatz, namely
\begin{equation} 
B^*=B\exp\left({-\frac{4g_\sigma^B\sigma}{M_N}}\right), \label{Bsigma}
\end{equation}
where $g_\sigma^B$ is an additional parameter and $B$ is the value of the bag 
constant in vacuum. In this work we use $g_\sigma^B=2.8$, which is the same 
as in Ref.~\cite{LTTWS}.

The quark-meson coupling constants $g_\sigma^q$ and $g_\omega = 3g_\omega^q$ are 
fitted to obtain the correct saturation properties of nuclear matter, $E_B \equiv 
E/A - M = \epsilon/\rho - M = -15.7$~MeV at $\rho~=\rho_0=~0.15$~fm$^{-3}$. 
We take the standard values for the meson masses, $m_\sigma = 550$~MeV and 
$m_\omega = 783$~MeV. 
We present results for three different values of the hard core, $r = 0.4$~fm, 
$0.5$~fm and $0.6$~fm, and for two values of bag radii, $R=0.6$~fm and $0.8$~fm. 
The pair $r=0.6$~fm, $R=0.6$~fm represents the situation that the size of the 
hard-core is the same as of the bag and is included for illustrative purposes. 

\begin{figure}[htb]
\protect
\epsfxsize=1.0\textwidth
\begin{center}
\leavevmode
\epsfbox{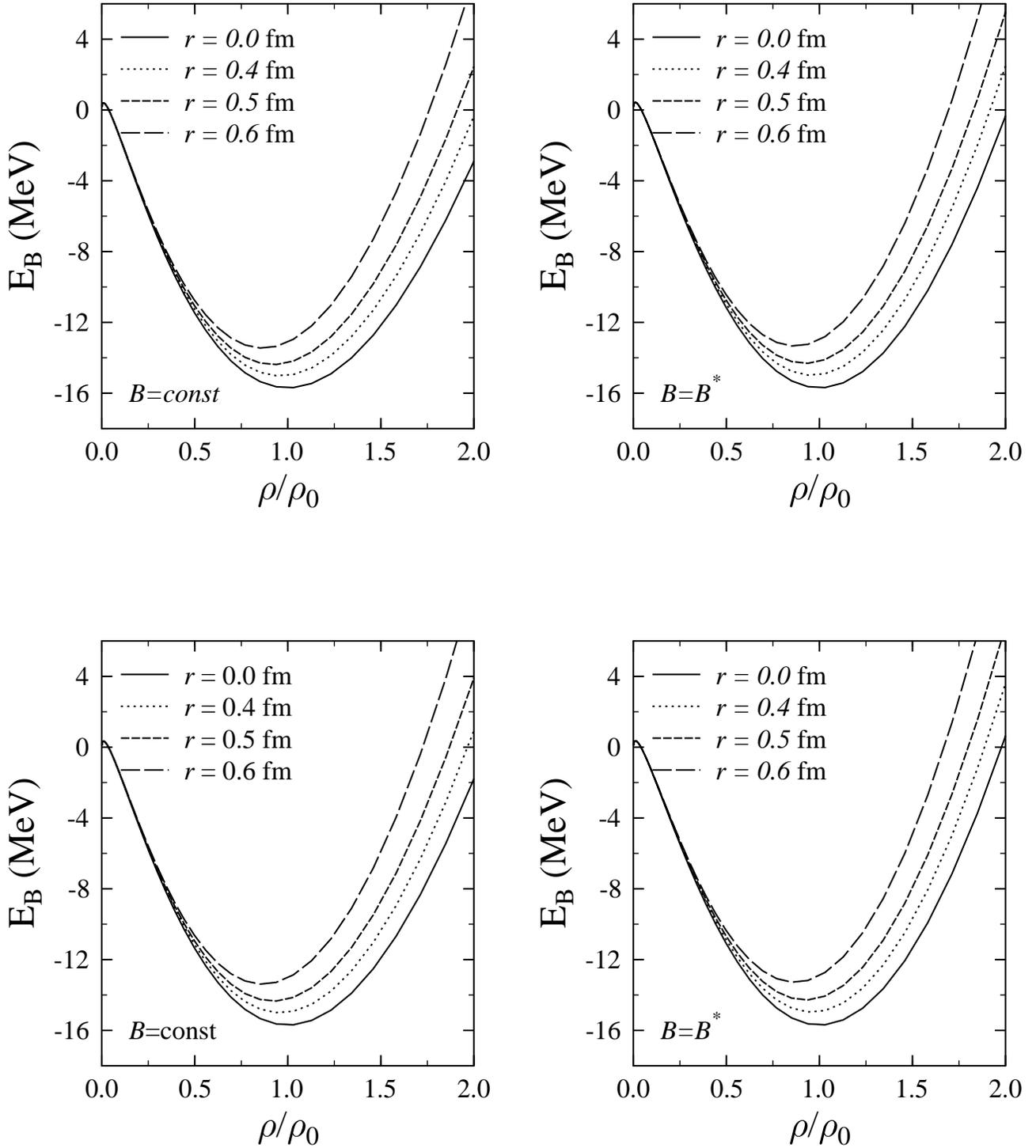}
\end{center}
\caption{The energy per nucleon of nuclear matter as a function of 
$\rho/\rho_0$ for different hard-core radii. All curves are for the same
set of quark-meson coupling constants. The upper pannel of the figure 
corresponds to 
$R=0.8$ fm and the lower one is for $R=0.6$ fm.}
\label{fig_comp}
\end{figure}

Initially, we investigate the effect of the excluded volume on the binding
energy per particle for the values of $r$ and $R$ mentioned above. The results 
for the different values of $r$ and~$R$ 
and shown in Fig.~\ref{fig_comp}, where we plot $E_B$ as a function of the 
nuclear density $\rho$. In this figure the coupling constants $g_\sigma^q$ and 
$g_\omega$ for a given $R$ are the same for the different values of~$r$.
As expected, the effect of an effective repulsion due to the 
hard core is clearly seen in this figure. The effect obviously increases as 
the size of hard-core increases. At the saturation density, the largest value 
of the effective repulsion is of 
the order of $4$~MeV. The effect is not as dramatic as one could expect. For
comparison with another repulsive effect, we mention that Fock 
terms~\cite{ffactors} give $5$~MeV repulsion for the binding energy. 

We now readjust the coupling constants $g_\sigma^q$ and $g_\omega$ such as
to obtain the correct saturation binding energy of nuclear matter for the 
different values of $r$ and $R$. Our aim is to investigate the changes on the 
properties of nuclear matter and in-medium nucleon properties due to the 
hard core. Tables~II and III present the values of the coupling constants 
and the ratios of in-medium to free-space bag radii $R^*/R$, nucleon masses 
$M^*/M$ and bag eigenvalues $x^*/x$. The Tables also show the changes in
the incompressibility for different hard core radii. The results are such
that  nucleon properties are not changed significantly, being at most 
at the level of $2\%$. The incompressibility is a little more sensitive 
than nucleon properties to the extra repulsion induced by the hard core,
but the increase is at most $120$~MeV. 

\begin{figure}[htb]
\protect
\epsfxsize=1.0\textwidth
\begin{center}
\leavevmode
\epsfbox{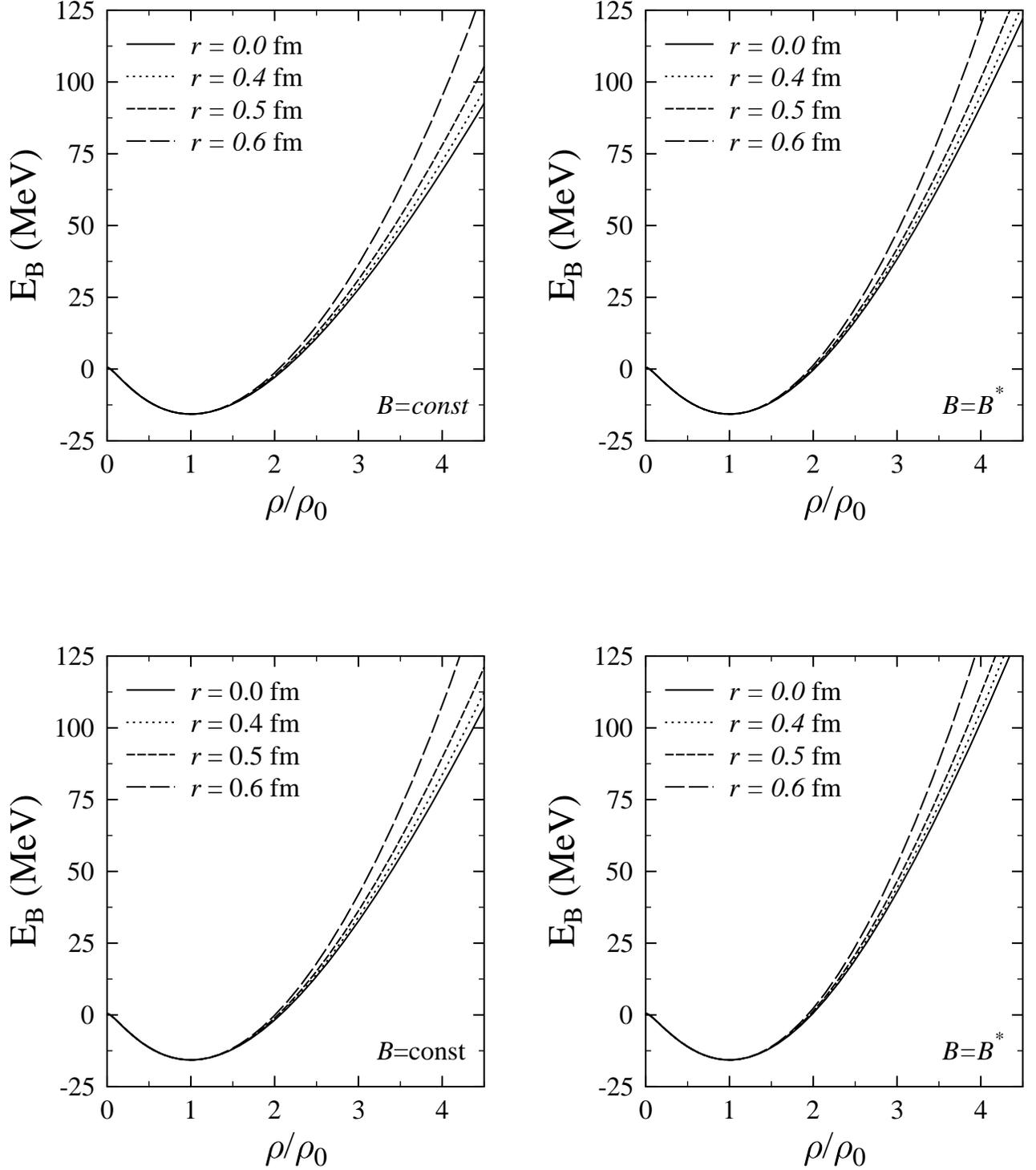}
\end{center}
\caption{The energy per nucleon of nuclear matter as a function of 
$\rho/\rho_0$ for different hard-core radii. The quark-meson coupling constants
are refitted such as to obtain the correct saturation point. The upper pannel 
of the figure corresponds to $R=0.8$ fm and the lower one is for $R=0.6$ fm.}
\label{EB-rs}
\end{figure}

\begin{figure}[htb]
\epsfxsize=1.0\textwidth
\begin{center}
\leavevmode
\epsfbox{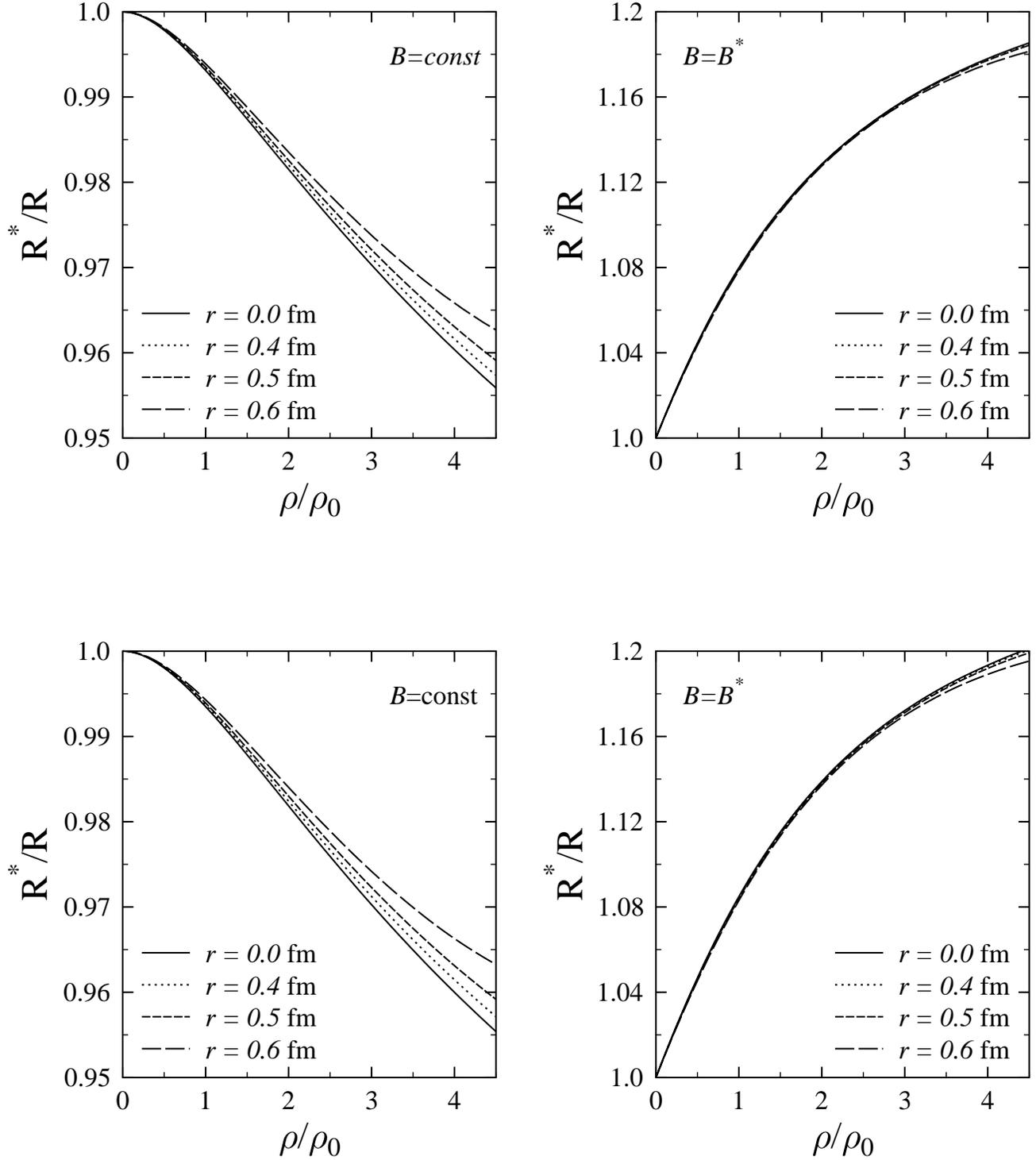}
\end{center}
\caption{The effective radius of the nucleon as a function of $\rho/\rho_0$ 
corresponding to the different hard-core radii. The upper pannel 
of the figure corresponds to $R=0.8$ fm and the lower one is for $R=0.6$ fm.}
\label{rstar}
\end{figure}

\begin{figure}[htb]
\epsfxsize=1.0\textwidth
\begin{center}
\leavevmode
\epsfbox{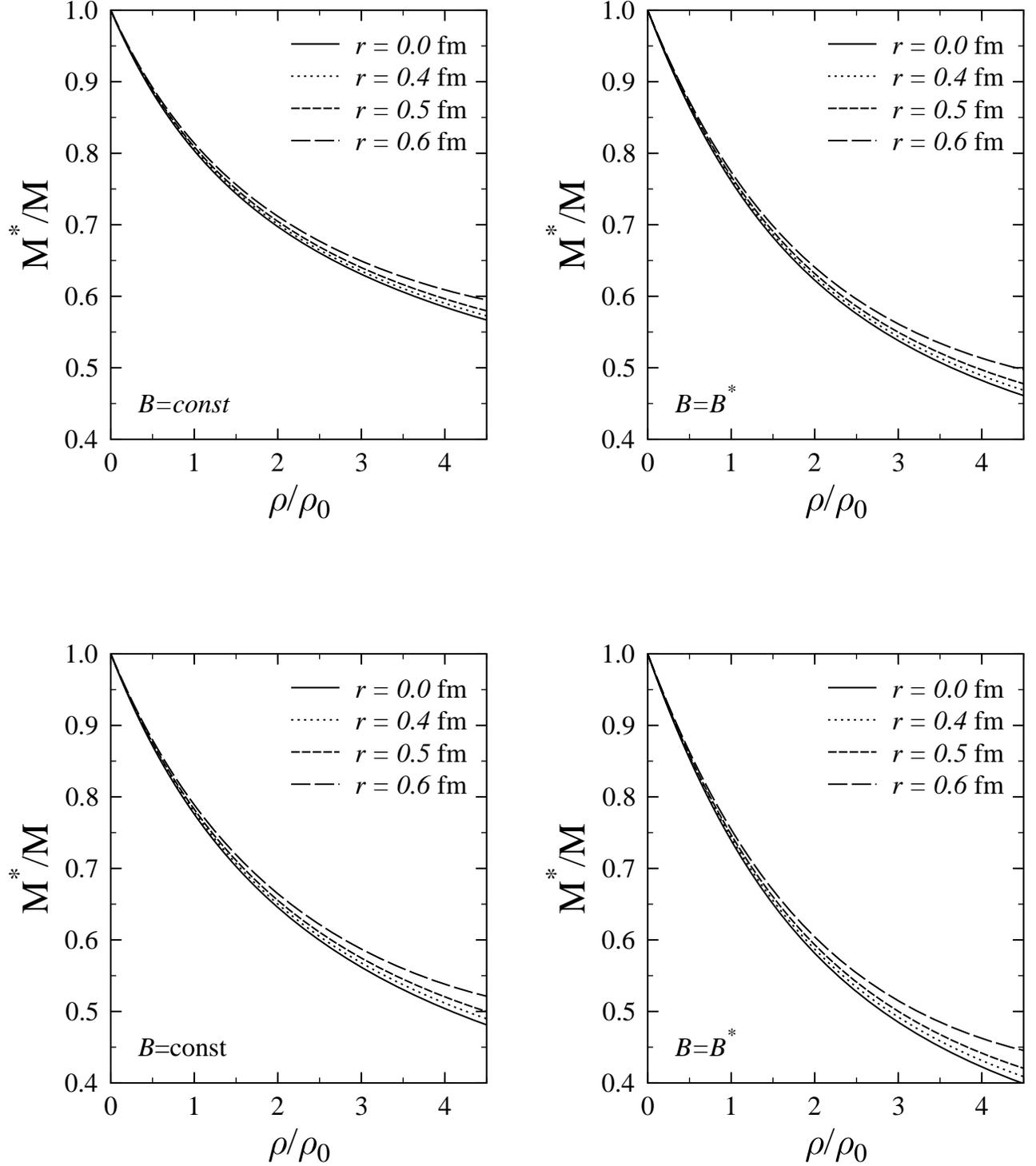}
\end{center}
\caption{The in-medium nucleon mass as a function of $\rho/\rho_0$ 
corresponding to the different hard-core radii. The upper pannel 
of the figure corresponds to $R=0.8$ fm and the lower one is for $R=0.6$ fm.}
\label{mstar}
\end{figure}

\begin{figure}[htb]
\epsfxsize=1.0\textwidth
\begin{center}
\leavevmode
\epsfbox{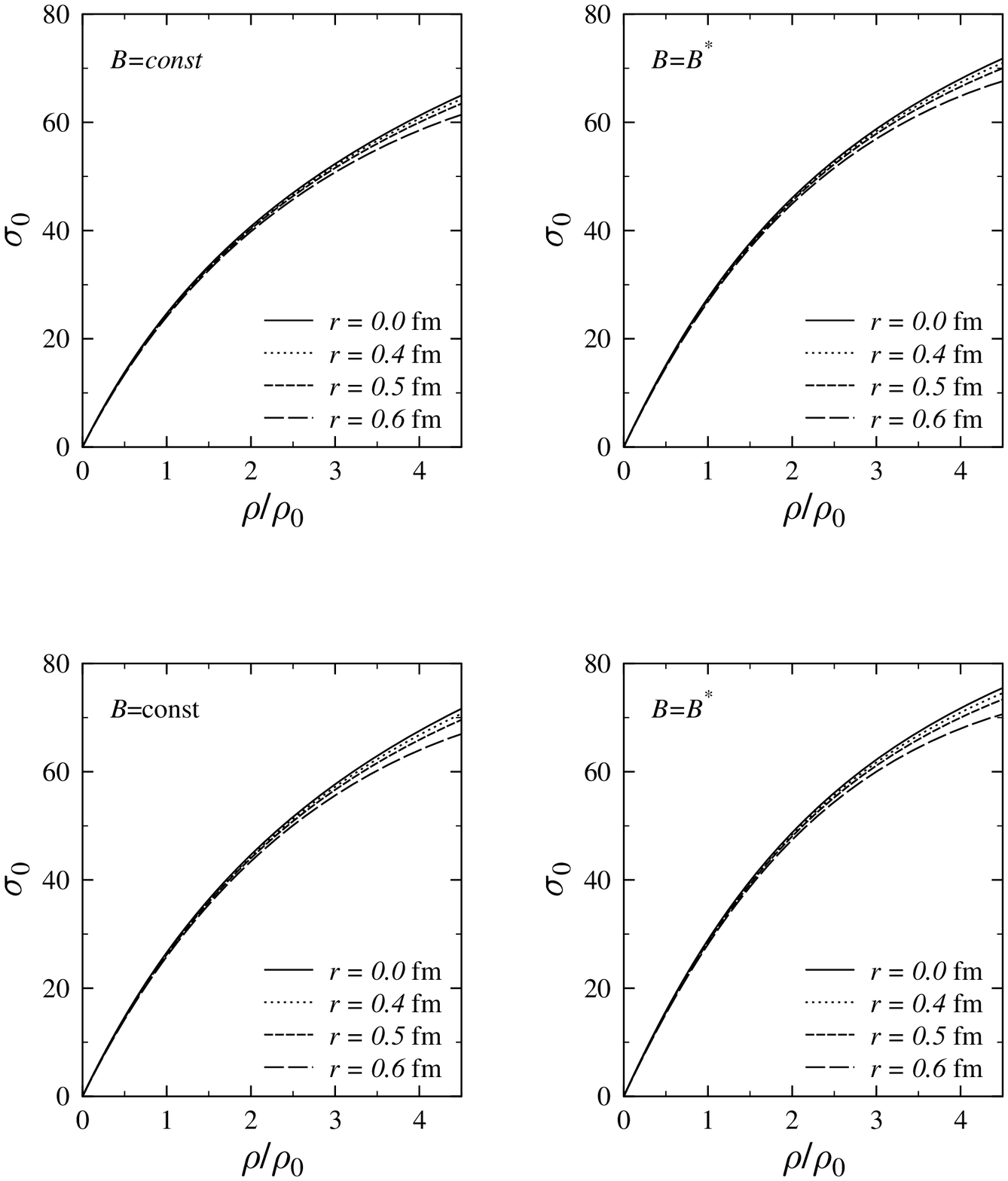}
\end{center}
\caption{The sigma field as a function of $\rho/\rho_0$ 
corresponding to the different hard-core radius. The upper pannel 
of the figure corresponds to $R=0.8$ fm and the lower one is for 
$R=0.6$ fm.}
\label{sigma}
\end{figure}

The effect of the hard core as a function of the nuclear density $\rho$ on
the binding energy is shown is Figure~\ref{EB-rs}. One notices that the 
equation of state becomes stiffer as the size of the hard core increases. 
The ratios $R^*/R$, $M^*/M$ and the $\sigma_0$ field as functions of $\rho$
are shown in Figures~\ref{rstar}, \ref{mstar} and \ref{sigma}, respectively.
As found previously, the in-medium bag radius decreases (increases) for a
constant (in-medium changed) bag parameter. Now, the change in the in-medium 
bag radius decreases as the hard core radius increases. This is clearly an
effect due to the fact that as the hard core radius increases, one has less 
attraction, and the bag properties change less. In Fig.~\ref{mstar}
one sees the interesting feature that as the in-medium nucleon mass 
increases, the binding energy curve is stiffer when volume corrections 
are included, contrary to the case without excluded volume. This is
again an effect of extra repulsion due to the hard core. That one gets less 
attraction as the hard core radius increases can be seen in 
Figure~\ref{sigma}, where we plot $\sigma_0$ as function of $\rho$ for
different combinations of $r$ and $R$. The less attraction is simply due to 
the factor $1+v_0\rho$ in the denominator in Eq.~(\ref{expl-sig}), which
increases as $v_0$ increases and makes the r.h.s. of Eq.~(\ref{expl-sig})
to contribute less to $\sigma_0$. 

\begin{figure}[htb]
\protect
\epsfxsize=1.0\textwidth
\begin{center}
\leavevmode
\epsfbox{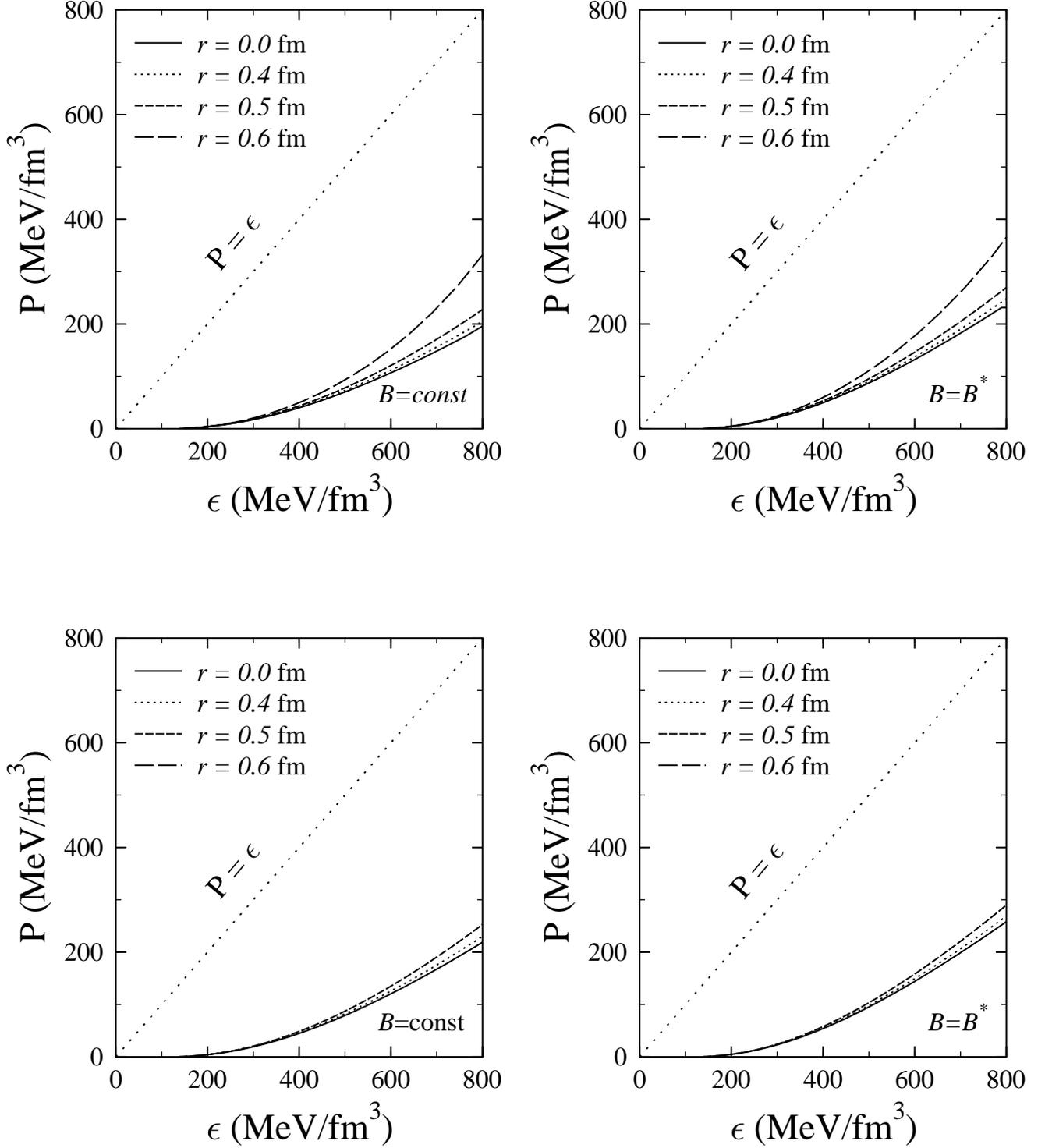}
\end{center}
\caption{The pressure of the nuclear matter as a function of the 
energy density corresponding to the different hard-core radii. The upper pannel 
of the figure corresponds to $R=0.8$ fm and the lower one is for $R=0.6$ fm.}
\label{press}
\end{figure}

To conclude this section,  we mention that for neutron stars, for instance, 
one is interested in the equation of state pressure $P$ versus energy 
$\epsilon$. One important 
question here is to check whether causality is respected by such an equation 
of state. Figure~\ref{press} presents $P$ versus $\epsilon$ for different 
values of $r$ and $R$. For 
comparison, the causal limit $P=\epsilon$ is also shown in the figure. 
Clearly seen is that all the cases studied here respect the causal
condition $\partial P/\partial\epsilon \le 1$, so that the speed of 
sound remains lower than the speed of light. This result is consistent 
with Ref.~\cite{causal}, where it was shown that for realistic situations
of temperatures below the QCD phase transition, which is believed to be 
of the order of $200$~MeV, the excluded volume prescription used 
here~\cite{Goren1,Goren2} does not lead to conflicts with 
causality. 

\section{Conclusions and Perspectives}
\label{concl}

In this paper we have incorporated excluded volume effects in the quark 
meson coupling model in a thermodynamically consistent manner. The excluded
volume simulates in a phenomenological way the short range hard-core repulsion 
of the nucleon-nucleon force, in the sense it does not allow nucleons to occupy
all space as they were point-like. The consequences for in-medium nucleon 
properties and saturation properties of nuclear matter due to the 
excluded volume effects have 
been investigated for different bag and hard core radii. The bag constant was
allowed to change in medium and differences with respect to a fixed bag
constant were studied. It was also shown that the prescription used does not 
lead to violations of causality.

We found that the excluded volume induces an effective repulsion that 
increases as the size of hard-core increases. The repulsion is at most 
$4$~MeV at the saturation density. In-medium nucleon properties, such as
bag radius and nucleon mass are not changed significantly, as compared to 
the changes when excluded volume effects are not taken into account. The 
changes are at most at the level of $2\%$. The incompressibility is a little 
more sensitive, but the increase is at most $120$~MeV. The excluded volume 
also induces the effect that the binding energy 
curve as a function of the nuclear density is stiffer as the in-medium 
nucleon mass increases. This feature is contrary to the case without excluded 
volume. It arises because of the extra repulsion due to the hard core that leads 
to a smaller sigma field and consequently to less attraction.

The formalism of the present paper can be extended in several ways. We
intend to incorporate taking into account of the Lorentz contraction
of the bags. As indicated by the authors of Ref. \cite{Goren3}, the 
effects of the Lorentz contraction is most important for light particles
like pions. Also possible extensions of the formalism presented here to
finite temperatures are currently under progress. As a final remark, it
should be clear that an excluded volume approach is by no means a complete
replacement of explicit calculations of short-range correlation effects,
such as through a Bethe-Goldstone--type of approach~\cite{GWW}. There is
one attempt to include short-range quark-quark correlations in the
QMC model~\cite{corr-QMC} and its further investigation in the context
of a Bethe-Goldstone approach is an interesting new direction 
that should be undertaken in the near future.

\section{Acknowledgments}
One of the author (PKP) would like to acknowledge
the IFT, S\~ao Paulo, for kind hospitality. This research has been supported 
in parts by CNPq and FAPESP (grant 99/08544-0).

\newpage

\begin{table}
\caption{Parameters used in the calculation.}\vspace{0.5in}

\begin{tabular}{cccccc}
$m_q$ (MeV) & $R$ (fm) & $B^{1/4} (MeV)$ & $z_0$ & $m_\sigma (MeV)$ 
& $m_\omega$ (MeV) \\ \hline
0 & 0.6 & 211.3 & 3.987 & 550 & 783 \\ 
0 & 0.8 & 170.3 & 3.273 & 550 & 783 \\ 
\end{tabular} 
\end{table}

\vspace{1.0in}

\begin{table}
\caption{quark-sigma and omega-nucleon are used for 
different cases in our calculation for $R=0.6$ fm. Effective nucleon radius, 
effective mass, bag eigen value and compressibility are given for different 
sets at the saturation density. Note that hard-core radius $r=0$ fm 
corresponds to normal QMC model}
\vspace{0.5in}

\begin{tabular}{cccccccc}
&Hard-core &~$g_s^q$~ &~ $g_\omega$~&~ $R^*/R$~&~$M^*/M_N$& ~$x^*/x$ & K\\
&radius (fm) & && && & MeV\\
\hline
&0 & 5.98 & 8.95 & 0.9934 & 0.7757 & 0.8659&257\\
&0.4& 5.93 & 8.81 & 0.9936 & 0.7789 & 0.8684&285\\
$B$=constant
&0.5& 5.87 & 8.66 & 0.9939 & 0.7824 & 0.8711&316\\
&0.6& 5.76 & 8.38 & 0.9942 & 0.7887 & 0.8759&372\\
\hline
&0 & 4.32 & 9.87 & 1.0849 & 0.7388 & 0.8882&268\\
&0.4& 4.26 & 9.72 & 1.0844 & 0.7426 & 0.8909&297\\
$B=B^*$
&0.5& 4.20 & 9.57 & 1.0839 & 0.7468 & 0.8938&330\\
&0.6& 4.09 & 9.29 & 1.0830 & 0.7543 & 0.8988&386\\
\end{tabular}                
\end{table}

\begin{table}
\caption{Same as Table II for $R=0.8$ fm}
\vspace{0.5in}

\begin{tabular}{cccccccc}
&Hard-core &~$g_s^q$~ &~ $g_\omega$~&~ $R^*/R$~&~$M^*/M_N$& ~$x^*/x$& K \\
&radius (fm) & && && & MeV\\
\hline
&0 & 5.74 & 8.19 & 0.9930 & 0.8034 & 0.8342&249\\
&0.4& 5.69 & 8.06 & 0.9932 & 0.8060 & 0.8371&277\\
$B$=constant
&0.5& 5.64 & 7.91 & 0.9935 & 0.8088 & 0.8404&307\\
&0.6& 5.54 & 7.64 & 0.9938 & 0.8139 & 0.8461&361\\
\hline
&0 & 4.14 & 9.34 & 1.0799 & 0.7609 & 0.8594&261\\
&0.4& 4.09 & 9.20 & 1.0795 & 0.7640 & 0.8627&290\\
$B=B^*$
&0.5& 4.03 & 9.05 & 1.0792 & 0.7675 & 0.8661&322\\
&0.6& 3.93 & 8.79 & 1.0785 & 0.7737 & 0.8723&378\\
\end{tabular}                
\end{table}

\begin{thebibliography}{99}
%
\bibitem{HP} H. Heiselberg and V. Pandharipande, Ann. Rev. Nucl. Part. Sci. 
{\bf 50}, 481 (2000).
%
\bibitem{HM} J.W. Harris and B. M\"uller, Ann. Rev. Nucl. Part. Sci. {\bf 46},
71 (1996).
%
\bibitem{FW} B. Friedman and V.R. Pandharipande, Nucl. Phys. {\bf A361}, 502
(1981). 
%
\bibitem{WFF} R.B. Wiringa, V. Fiks, and A. Fabrocini, Phys. Rev. C {\bf 38},
1010 (1988).
%
\bibitem{WalSer} J.D. Walecka, Ann. of Phys. {\bf 83}, 491 (1974); 
B.D. Serot and J.D. Walecka, Int. J. Mod. Phys. E {\bf 6}, 
515 (1997).
%
\bibitem{Eff} R.J. Furnstahl, B.D. Serot and H.B. Tang, Nucl. Phys 
{\bf A615}, 441 (1997); H. Mueller and B.D. Serot, Nucl. Phys. {\bf A606}, 
508 (1996).
%
\bibitem{bagEOS} J.C. Collins and M.J. Perry, Phys. Rev. Lett. {\bf 34},
1353 (1975); G. Baym and S.A. Chin, Phys. Lett. B {\bf 62}, 241 (1976);
B.D. Keister and L.S. Kisslinger, Phys. Lett. B {\bf 64}, 117 (1976);
G. Chapline and M. Nauenberg, Phys. Rev. D {\bf 16}, 450 (1977); 
B. Freedman and L. McLerran, Phys. Rev. D {\bf 17}, 1109 (1978).
%
\bibitem{MITbag} A. Chodos, R.L. Jaffe, K. Johnson, C.B. Thorn, and
V. Weisskopf, Phys. Rev. D {\bf 9}, (1974) 3471; A. Chodos, R.L. Jaffe, 
K. Johnson, and C.B. Thorn, Phys. Rev. D {\bf 10}, 2599 (1974).   
%
\bibitem{FT} D. Hadjimichef, G. Krein, S. Szpigel, and J.S. da 
Viega, Ann. Phys. (N.Y.) {\bf 268}, 105 (1998).
%
\bibitem{guichon} P. A. M. Guichon, Phys. Lett. {\bf B 200}, 235 (1988). 
%
\bibitem{ST} K. Saito and A.W. Thomas, Phys. Lett. B {\bf 327}, 9 (1994);
{\bf 335}, 17 (1994); {\bf 363}, 157 (1995); Phys. Rev. C {\bf 52}, 2789 
(1995); P.A.M. Guichon, K. Saito, E. Rodionov, and A.W. Thomas, Nucl. Phys. 
{\bf A601} 349 (1996);  K. Saito, K. Tsushima, and A.W. Thomas, Nucl. Phys. 
{\bf A609}, 339 (1996); Phys. Rev. C {\bf 55}, 2637 (1997); Phys. Lett. B 
{\bf 406}, 287 (1997); K. Tsushima, K. Saito, J. Haidenbauer, and
A. W. Thomas, Nucl. Phys. A {\bf 630}, 691 (1998).
%
\bibitem{earlier} T. Frederico, B.V. Carlson, R.A. Rego, and M.S. Hussein,
J. Phys. G {\bf 15}, 397 (1989); S.~Fleck, W. Bentz, K. Shimizu, K. Yazaki,
Nucl. Phys. A {\bf 510}, 731 (1990); E. Naar, M.C. Birse, J. Phys. G {\bf 19},
555 (1993).
%
\bibitem{ffactors} G. Krein, A.W. Thomas, and K. Tsushima,
Nucl. Phys. A {\bf 650}, 313 (1999); M.E. Bracco, G. Krein, and 
M. Nielsen, Phys. Lett. B {\bf 432} 258 (1998).
% 
\bibitem{bstar} X. Jin and B.K. Jennings, Phys. Lett. B 374, 13 
(1996); Phys. Rev. C {\bf 54}, 1427 (1996).
%
\bibitem{recent} P. G. Blunden and G.A. Miller, Phys. Rev. C 
{\bf 54}, 359 (1996);  G. Krein, D.P. Menezes, M. Nielsen, 
and  C. Providencia, Nucl. Phys. {\bf A674}, 125 (2000); 
N. Barnea and T.S. Walhout, Nucl. Phys. {\bf A677}, 367 (2000);
H. Shen and H. Toki, Phys. Rev. C {\bf 61}, 045205 (2000).
%
\bibitem{stars} S. Pal, M. Hanauske, I. Zakout, H. St\"ocker, and 
W. Greiner, Phys. Rev. C {\bf 60}, 015802 (1999).
%
\bibitem{LTTWS} D.H. Lu, K. Tsushima, A.W. Thomas, A.G.
Williams and K. Saito, Nucl. Phys. {\bf A634}, 443 (1998).
%
\bibitem{temp} P.K. Panda, A. Mishra, J.M. Eisenberg, W. Greiner,
Phys. Rev. C {\bf 56}, 3134 (1997); I. Zakout, H.R. Jaqaman, S. Pal,
H. St\"ocker, and W. Greiner, Phys. Rev. C {\bf 61}, 055208 (2000).
%
\bibitem{qmcstar} P.K. Panda, R. Sahu, C. Das, 
Phys. Rev. C {\bf 60}, 38801 (1999).
%
\bibitem{GWW} L.C. Gomes, J.D. Walecka and V.F. Weisskopf, Ann. Phys.
(N.Y.) {\bf 3}, 241 (1958); J.D. Walecka and L.C. Gomes, Ann. da Acad. 
Brasileira de Ci\^encias {\bf 39}, 361 (1967). 
%
\bibitem{Goren1} M.I. Gorenstein, V.K. Petrov, and G.M. Zinovjev,
Phys. Lett. B {\bf 106}, 327 (1981).
%
\bibitem{Goren2} D.H. Rischke, M.I. Gorenstein, H. St\"ocker and 
W. Greiner, Z. Phys. C {\bf 51}, 485 (1991); B.Q. Ma, Q.R. Zhang, 
D.H. Rischke and W. Greiner, Phys. Lett. B {\bf 315}, 29 (1993).
%
\bibitem{goldman}T. Goldman and G. J. Stephenson, Jr., Phys. Lett. B {\bf 146},
143 (1984);
T. Goldman, G. J. Stephenson, Jr., and K. E. Shmidt, Nucl. Phys. A {\bf 481},
621 (1988);
Fang Wang, Guang-han Wu, Teng, L.J. , and T. Goldman,
Phys. Rev. Lett. {\bf 69}, 2901 (1992);
T. Goldman, K. Maltmann, and G. J. Stephenson, Jr., Phys. Lett. B {\bf 324},
1 (1994);
Guang-han, Jia-Lun Ping, Li-jian Teng, Fan Wang, and T. Goldman,
Nucl. Phys. A {\bf 673}, 279 (2000).
%
\bibitem{Goren3} K.A. Bugaev, M.I. Gorenstein, H. St\"ocker, and
W. Greiner, Phys. Lett. B {\bf 485}, 121 (2000).
%
\bibitem{causal} N. Prasad, K.K. Singh, and C.P. Singh, Phys.
Rev. C {\bf 62}, 037903 (2000).
%
\bibitem{corr-QMC} K. Saito and K. Tsushima, Prog. Theor.
Phys. (Kyoto) {\bf 105}, 373 (2001).
\end{thebibliography}
\end{document}